# Paradyn BlendOpt™

Making Mining More Efficient

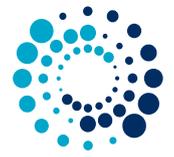

Whitepaper

**Coal Blending: Business Value, Analysis, and Optimization**

Dr James Whitacre, Director, Paradyn  www.paradynsystems.com
Dr Antony Iorio, Director, Paradyn
Sven Schellenberg, Director, Paradyn

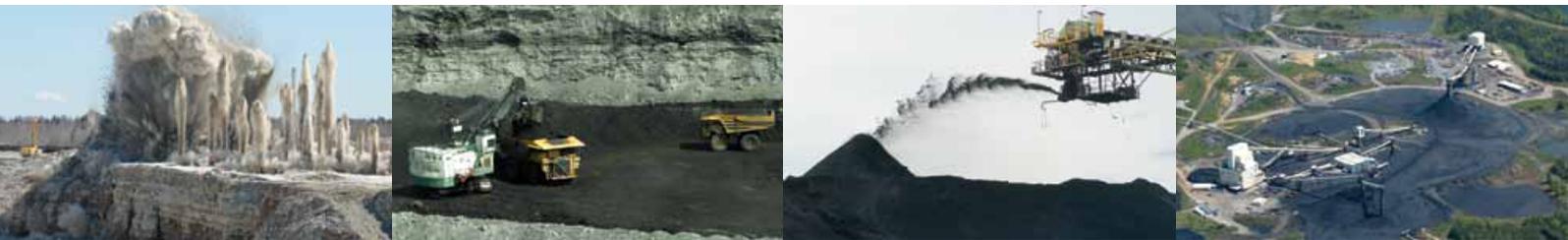

Coal blending is a critically important process in the coal mining industry as it directly influences the number of product tonnes and the total revenue generated by a mine site.  Coal blending represents a challenging and complex problem with numerous blending possibilities, multiple constraints and competing objectives. At many mine sites, blending decisions are made using heuristics that have been developed through experience or made by using computer assisted control algorithms or linear programming.  While current blending procedures have achieved profitable outcomes in the past, they often result in a sub-optimal utilization of high quality coal. This sub-optimality has a considerable negative impact on mine site productivity as it can reduce the amount of lower quality ROM that is blended and sold. This article reviews the coal blending problem and discusses some of the difficult trade-offs and challenges that arise in trying to address this problem. We highlight some of the risks from making simplifying assumptions and the limitations of current software optimization systems.  We conclude by explaining how the mining industry would significantly benefit from research and development into optimization algorithms and technologies that are better able to combine computer optimization algorithm capabilities with the important insights of engineers and quality control specialists.

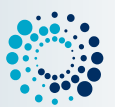

# Background

Coal blending is a common practice for achieving quality attributes needed in a specific application, e.g. steam generation, coking. The quality attributes of greatest concern will vary from one mine to another and depend on Run Of Mine (ROM; i.e. raw, unprocessed coal) quality variation between plies or seams and the product's application. For thermal coal, properties such as ash, volatile matter, total sulfur, moisture, Hargrove Grindability Index, and specific energy are significant to the product. For coking coals, additional attributes can include crucible swelling number, fluidity, and RoMax.

## Blending Methodology

Blending can take place in several locations across a supply chain with the location of blending sometimes being important. For instance, blending before the Coal Handling and Preparation Plant (CHPP, i.e. wash plant) is sometimes needed to improve washability and increase CHPP production rates. The act of blending typically occurs through stockpile stacking however it may also take place within a vessel's hatch during ship loading. Stacking methodology (e.g. Chevron, Windrow, Cone Shell, Strata) influences homogeneity of the final blended material and subsequently influences sampling consistency and the risk of failing quality validation tests.

**Blending Points within the Coal Chain**

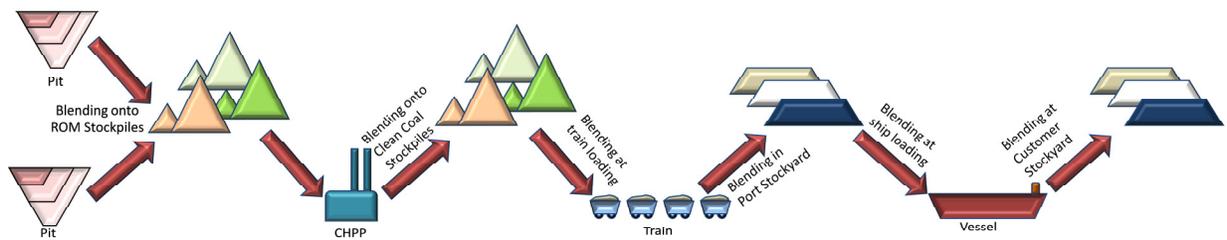

## Business Impact

Without having tested alternative blending software technologies, many coal mines are unaware of the impact of these decisions on mine productivity. This is particularly true for coal mines that sell multiple products. Each product has its own unique set of quality ranges that must be satisfied based on customer contracts. Within each quality range, there can also be a desirable quality target that reflects ideal product conditions for a customer. Customers and mine sites will therefore sometimes negotiate a penalty or bonus that modifies the final sale value of a product based on a quality attribute's relative position within the contractually required quality range, e.g. its degree of deviation from a desired quality target. Each product thus may have a different base sale price, different financial penalties and bonuses, and considerably different quality requirements.

Changes to a product's blend can impact the sale value of the product, the total amount of in spec tonnes of a product, and the amount of ROM available for creating other products. Consequently, blending decisions directly influence the number of product tonnes and the total revenue of a mine site.

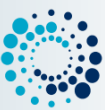

# Blend optimization.

Blending is a nonlinear combinatorial optimization problem where the objective is typically to maximize revenue, Net Present Value (NPV), or monthly product tonnage targets. Important and closely related Key Performance Indicators (KPIs) include total mine site productivity (e.g. total sold tonnes per month), product quality consistency, and product reliability (e.g. consistent availability of a product from month to month).

**Interdependencies within the Coal Blending Problem**

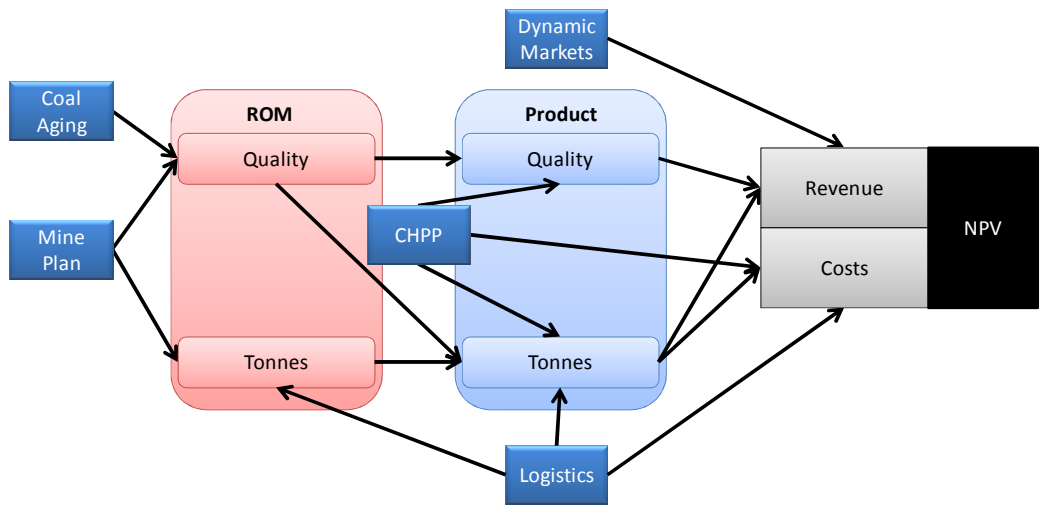

Products are generally created by blending together several different ROM types (many to 1 mapping) while each ROM type can contribute to multiple products (1 to many mapping). These mathematical mapping features can result in a very large number of unique blend possibilities. For example, consider a mine site with five different ROM types that each can be blended in 1000 tonne allotments to create one of two products that each have a desired tonnage target of 100KT. For this simplified blending problem there exists approximately 18 trillion blending possibilities.

## Quality versus Tonnage

A ROM type can often be added to several different products however the ROM's influence on each product's sale value will depend on the qualities of other available ROMS and the required quality ranges of the product. Once contractual tonnage obligations have been satisfied, it is common for blending heuristics[1] to be used that focus attention on one particular product more than others. For instance, it is sometimes assumed that blending that maximizes tonnage of the most profitable product will correspond with the best business decisions for the mine. This is similar to the assumption that the best business decision will be that corresponding with the best average revenue per product tonne. Alternatively, it is sometimes assumed that blending changes that maximize total productive tonnes up to supply chain capacity constraints will correspond with the best business decisions. A variety of other heuristics have been developed that are tailored to a specific mine site and aim to strike a balance between maximizing total productive tonnes and maximizing the average sale value per tonne.

---

[1] A heuristic is a method - developed through experience, judgment, and training - that is likely to yield a good solution to a problem but that is not guaranteed, or even likely, to produce an optimal outcome.

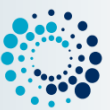

In practice however, the best business outcome cannot be reliably characterized by any single set of heuristics and instead depends on a number of factors that can change at a mine site over time. For example, consider a situation where it is necessary to include 20% of ROM type A to sweeten poor quality ROMs within a high volatile coking coal product, while 45% of ROM type A is needed to sweeten poor quality ROMs in order to create a low volatile coking coal product. The optimal decision in this case will depend on the difference in price between the high volatile and low volatile products, the abundance of poorer quality ROMs, the distribution of ROM quality attributes, the availability of sufficient production capacity, processing costs, and contractual commitments.

Similar difficulties arise when considering whether a product should be blended to achieve a customer's ideal quality targets. On the one hand, a customer may agree to pay a higher price for products that are closer to a particular quality target. On the other hand, the higher quality ROM needed to achieve this target might instead be used to create more tonnage of other products. There are no generic rules of thumb that will be best in addressing these trade-offs due to price fluctuations and the dynamic availability of ROM types.

### Dynamic Markets

Spot and negotiated sale prices for products change over time in response to changing market conditions. When specific products have strong sale price growth forecasts, situations can arise where it is more profitable for the business to reserve tonnage of selected ROMs for future blends. Ideally, decisions to delay profits should evaluate the impact on NPV or Economic Rate of Return (ERR) in order to be justifiable to shareholders. In addition, the quality attributes of coal can degrade as it sits on a stockpile and is exposed to rain and oxidizing conditions. Consequently, the "ROM excavation birthday" must also be considered when making blending decisions.

Although revenue and NPV are obviously important to blending decisions, it is also strategically important that contractual commitments are satisfied. Some customers may have long standing large contracts for specific products and the continuation of these commitments mitigate real uncertainty in business valuation, e.g. in discounted cash flow analysis, which is important to shareholders. As a result, it is often (incorrectly) assumed that blending decisions should firstly prioritize contractually binding tonnage targets regardless of whether these decisions correspond with the most profitable outcome.

### CHPP Ash versus Yield

ROM is sometimes washed to remove ash, e.g. by gravity separation. The density cut-point setting of a wash plant's heavy medium circuits can be adjusted to remove additional ash from the feed ROM, however this lowers production yield and results in fewer saleable product tonnes. Each ROM will have its own unique relationship between initial (feed) ash, product ash, and product yield (i.e. the ash-yield curve) and the optimal wash plant operating conditions will depend on product requirements as well as the quality attributes of available ROM.

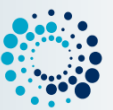

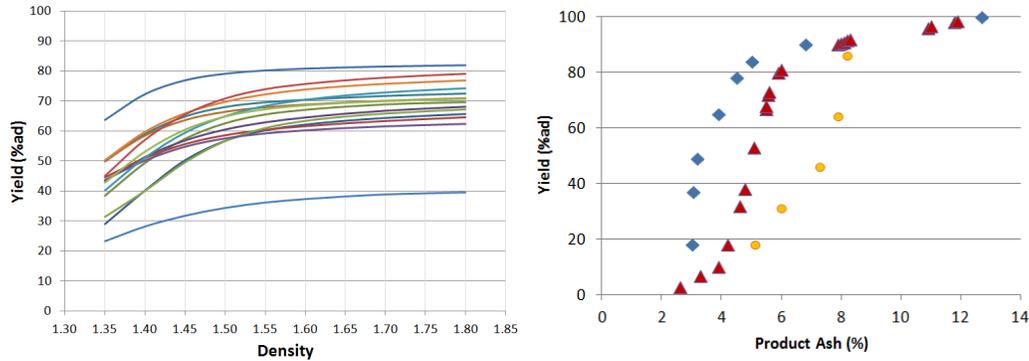

A wash plant manager will often aim to achieve a high and consistent operating rate in order to maximize wash plant utilization. This decision is partly justified by the fact that major wash plant operating costs (e.g. human resources, electricity, reagents) are only marginally influenced by production rates, and thus lower flow rates significantly increase the per tonne cost of washing coal. However, because of the complex relationships between tonnage, quality, and sale value of different products, maximum wash plant utilization rarely corresponds with maximum NPV.

Similarly, there are other sunk costs within the supply chain related to haulage, rail, and shipping contracts in which underutilization also results in higher costs per product tonne and it is often incorrectly assumed that the maximum utilization of these resources will correlate with maximum NPV.

### Logistical Constraints

There are a number of operational constraints and resource constraints that should be taken into account when making blending decisions. ROM is sometimes sourced from different pits with significantly different distances to the wash plant. Haulage fleet availability can therefore limit the amount of ROM that can be sourced from geographically distant pits. These haulage constraints can be partially mitigated, but not eliminated, by moving ROM to staging stockyards that are closer to the CHPP during time of excess haulage availability, however this also comes at a rehandling cost and can sometimes result in small material losses.

There are additional logistical constraints that can restrict the number of ROM types within a blend such as the location of truck loaders. Loader location and truck size also introduce practical limits on the minimum tonnage of a single ROM type that can be added to a blend.

## Blend Optimization Software

### Challenges in Manual Blending

The variety of blending possibilities is too large and complex for any engineer to construct an optimal solution. Instead blending decisions are made in the same way that complex scheduling decisions are handled within manufacturing and other industries, namely through satisficing and heuristics [1]. Satisficing - a term coined by Nobel Laureate Herbert Simon- refers to decision making that aims to achieve a feasible outcome with satisfactory performance and operates under the assumption that uncertainties and problem complexity will prevent optimality [2] [3] [4]. In blending problems, heuristics or rules of thumb are often developed through experimentation and experience that guide the creation of

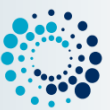

"good enough" blends. Unfortunately these heuristics can turn into very poor decisions when there are changes to the relative prices of different products.

## Software-Assisted Blending

The variety of variables, constraints, trade-offs, and uncertainties are too complex for any person to solve. Optimization software can be used to support these decisions and some mining software vendors provide blending capabilities. The optimization algorithms that have been designed in these systems often make simplifying assumptions in order to eliminate problem complexities (e.g. nonlinearities in objectives and constraints) so that the problem can be solved using linear programming or a similarly simplified optimization algorithm. Unfortunately, software simplifications can generate solutions that are sometimes not practically feasible and the solutions themselves are sometimes of lower quality than those generated by a skilled engineer. Case studies within other industries have similarly discovered that skilled schedulers and planners can often generate better schedules than optimization software, even for complex scheduling problems [5] [6].

If blending problems are purportedly too difficult for humans to optimize, then why can skilled engineers sometimes generate blends that are more desirable to the business? Resolving this paradox requires an appreciation for the importance of mine-site specific domain knowledge that is known by engineers but absent from the optimization model.

## Current Limitations of Blend Optimization Software

In practice, each mine site has unique characteristics in their blending problem that are understood by engineers but cannot be accounted for by reconfiguring the blending optimization algorithms. This may include any number of the topics already discussed in this article such as logistical constraints, wash plant considerations, contractual commitments, but also other issues not covered here such as the knowledge of losses/dilutions (e.g. from blast design, selective loading, transport), delay risks when implementing the current schedule, sociopolitical factors within the organization, and underlying tacit knowledge. In other words, blending requirements at different mine sites would appear to be too diverse for a single optimization technology to be able to effectively solve them all.

Importantly, there are proven theories in optimization research (cf No Free Lunch Theorems, NFL, for Optimization, [7] [8]) which actually guarantee that no general purpose blending optimizer can exist that will be perfectly designed to address every coal mine site's blending requirements![2] This well-tested theory is known throughout the optimization research community but it is not readily shared between software vendors and prospective customers.

## Recommendations

The implications of NFL to blend optimization software are considerable. First, if blend optimization software is to be broadly effective, it must facilitate user driven adjustments to blends. In some cases a planner will want to change specific attributes of a production plan but may not want to go through the dozens of tedious adjustments in order to explore and discover the necessary blending adjustments. For example, a planner might want to lower ash by an additional 2%, increase product tonnage by 10KT for a given month, or replace one ROM type with others in order to reserve that ROM for other products. If optimization engines were redesigned to support user-guided optimization, the expertise of the planner could be reflected in blending outcomes without requiring large numbers of tedious manual adjustments.

---

[2] Another important implication of NFL is that the more general purpose an optimizer claims to be, the less relevant its results are likely to be to your specific problem.

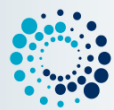

### Blend Analytics

Domain expertise is most effective when supported by the information needed to make sound judgments. Blend analysis is the process of evaluating the consequences of blending options within a schedule. For example, analysis can take place to determine a blend's impact on product quality, projected revenue, or the time criticality of scheduled mining activities for achieving production targets. An analytics platform that provided transparency across the myriad factors impacting blending decisions would help to better inform a planner of the business outcomes associated with different blending options. Combined with the user-guided optimization technologies just discussed, blending analytics could enable a faster and more thorough assessment of different blend possibilities and ultimately a more intelligent utilization of a mine site's most valuable ROM.

## Conclusions

Coal blending is a complex problem with a significant impact on mine site revenue and productivity. Blend optimization has thus far been achieved using algorithms that make simplifying mathematical assumptions, with results that are sometimes worse than manual blending decisions. This article has reviewed and discussed some of the limitations of optimization technologies currently available to the mining industry. Based on our experience in real-world optimization and coal blending in particular, this article proposes guidelines for improving the business value of blending optimization software by combining the computational power of optimization algorithms with the unique insights, experience, and knowledge of the engineer. User-guided optimization is not a new concept and there are numerous case studies demonstrating its significant positive effect on planning outcomes (e.g. see [9] [10] [11]), however the benefits of user-guided search have yet to be recognized by software vendors in the mining industry.

## Bibliography

[1] Y. Boasson, "Human aspects of scheduling: a case study," MIT, 2007.
[2] H. A. Simon, *A Behavioral Model of Rational Choice*. Santa Monica: Rand Corp, 1953.
[3] J. G. March, "Exploration and exploitation in organizational learning," *Organization Science*, vol. 2, no. 1, pp. 71–87, 1991.
[4] D. A. Levinthal and J. G. March, "The myopia of learning," *Strategic Management Journal*, vol. 14, no. 8, pp. 95–112, 1993.
[5] S. G. Nakamura N., "An experimental study of human decision-making in computer-based scheduling of flexible manufacturing system.," *International Journal of Production Research*, vol. 26, no. 4, p. 567, 1988.
[6] S. Crawford and B. L. et al. MacCarthy, "Investigating the work of industrial Schedulers through Field Study," *Cognition, Technology & Work*, vol. 1, no. 2, pp. 63–77, 1999.
[7] D. H. Wolpert and W. G. Macready, "No free lunch theorems for optimization," *IEEE Transactions on Evolutionary Computation*, vol. 1, no. 1, pp. 67–82, 1997.
[8] Y. C. Ho and D. L. Pepyne, "Simple explanation of the no-free-lunch theorem and its implications," *Journal of Optimization Theory and Applications*, vol. 115, no. 3, pp. 549–570, 2002.
[9] D. H. Baek and S. Y. et al Oh, "A visualized human-computer interactive approach to job shop scheduling," *International Journal of Computer Integrated Manufacturing*, vol. 12, no. 1, pp. 75–83, 1999.
[10] M. Chimani, N. Lesh, M. Mitzenmacher, C. Sidner, A. Case, and L. I. Optimization, "A Case Study in Large-Scale Interactive Optimization," *Elements*, 2004.
[11] G. W. Klau, N. Lesh, J. Marks, and M. Mitzenmacher, "Human-guided search," *Journal of Heuristics*, vol. 16, no. 3, pp. 289–310, 2009.

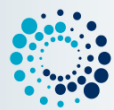